\documentclass[conference]{IEEEtran}
\IEEEoverridecommandlockouts
\usepackage{cite}
\usepackage{amsmath,amssymb,amsfonts}
\usepackage{algorithmic}
\usepackage{graphicx}
\usepackage{textcomp}
\usepackage{xcolor}

\usepackage[utf8]{inputenc}
\usepackage[T1]{fontenc}
\usepackage{svg}
\usepackage{hyperref}
\usepackage{fancyhdr}
\usepackage{tocloft}
\usepackage{lscape}
\usepackage{pdflscape}
\usepackage{algorithm}

\usepackage{booktabs}    
\usepackage{multirow}    
\usepackage{siunitx}     
\usepackage{caption}     
\usepackage{makecell} 
\usepackage{tabularx}
\usepackage{stfloats}
\usepackage{pifont}      
\newcommand{\cmark}{\ding{51}}  
\newcommand{\xmark}{\ding{55}}  

\newcolumntype{Y}{>{\centering\arraybackslash}X}

\def\BibTeX{{\rm B\kern-.05em{\sc i\kern-.025em b}\kern-.08em
    T\kern-.1667em\lower.7ex\hbox{E}\kern-.125emX}}
\begin{document}

\title{A Multi-Armed Bandit–Based Participant Selection Method for Federated Recommendation Systems\\
}

\author{\IEEEauthorblockN{Jintao Liu\textsuperscript{*}}
\IEEEauthorblockA{\textit{Faculty of Information Technology} \\
\textit{Monash University}\\
Melbourne, Australia \\
jintaoliu@protonmail.com}
\and
\IEEEauthorblockN{Mohammad Goudarzi\textsuperscript{*}}
\IEEEauthorblockA{\textit{Faculty of Information Technology} \\
\textit{Monash University}\\
Melbourne, Australia \\
mohammad.goudarzi@monash.edu}
\and
\IEEEauthorblockN{Adel Nadjaran Toosi}
\IEEEauthorblockA{\textit{School of Computing \& Information Systems} \\
\textit{The University of Melbourne}\\
Melbourne, Australia \\
adel.toosi@unimelb.edu.au}
\thanks{\textsuperscript{*}Corresponding authors: Mohammad Goudarzi and Jintao Liu.}
}

\maketitle

\begin{abstract}
Federated Recommendation Systems (FRS) enable privacy-preserving model training by keeping user data on edge devices. However, the practical deployment of FRS in Edge-Cloud environments faces significant challenges due to system and statistical heterogeneity. Existing FRS participant selection strategies struggle to dynamically balance the trade-off between model convergence speed and recommendation quality in such volatile environments. To address this, we formulate the FRS participant selection problem as a normalized utility cost addressing the model quality and system efficiency. Next, we propose a dynamic participant selection framework incorporating a Multi-Armed Bandit (MAB)-based solver for multimodal FRS. We design a client-utility function that jointly evaluates historical Client Performance Reputation, data quality, and real-time system latency. By leveraging an Upper Confidence Bound strategy, our framework effectively balances the exploration of under-sampled clients with the exploitation of high-performing ones. We validate the proposed approach on a realistic edge-cloud testbed implementation using a multimodal movie-recommendation task. Experimental results demonstrate that our MAB-driven approach outperforms other baselines across eight different data-skew scenarios. Specifically, it improves training efficiency by 32--50\% while improving model quality metrics such as Recall@50 by up to around 5\%.
\end{abstract}


\section{Introduction}
With the ongoing information explosion, users are provided with an exponentially growing scale of digital content, exemplified by Amazon’s 2024 catalog exceeding 353 million products \cite{coppola_2025_amazon} and TikTok’s 8.6 billion video uploads \cite{ceci2024tiktok}. At this scale, manual discovery of user interests becomes impractical, necessitating the adoption of recommendation systems (RS) to model user preferences and facilitate decision-making. 

Traditionally, RS have followed a cloud-centric paradigm, where service providers collect interactions, profiles, and auxiliary data to train models on central servers \cite{10925927}. Early methods relied on matrix factorization (MF) to decompose interaction matrices into latent factors; however, they were not well-suited for modeling complex user behaviors. The advent of deep neural networks has enabled capturing nonlinear patterns more effectively, further exploiting higher-order user–item relationships and yielding significant accuracy gains \cite{10144391}. 

Although cloud-based RS achieve high accuracy by leveraging large-scale interaction data, they struggle to accommodate the growing diversity and volume of multimodal user signals (i.e., textual reviews, images, audio clips, and videos) without incurring latency, bandwidth consumption, and privacy risks~\cite{10.1145/3695461}. Furthermore, despite impressive progress, centralized RS with cloud-based approaches inherently demand the continuous collection and storage of sensitive personal data. Such concentration of information not only heightens the exposure to large-scale breaches but also conflicts with an evolving landscape of privacy legislation \cite{GUENDOUZI2023103714}. Regulatory frameworks like the European Union’s General Data Protection Regulation (GDPR) \cite{voigt2017gdpr} and China’s Personal Information Protection Law (PIPL) \cite{calzada2022pipl} underscore a global shift toward user data sovereignty, enforcing strict controls on how personal information may be gathered, processed, and transferred. These mandates present a profound challenge for centralized RS. 

Edge–cloud architectures address these challenges by partitioning the recommendation pipeline, where lightweight models deployed at the network edge extract features and perform preliminary inference on raw multimodal inputs, while the cloud aggregates the full spectrum of user data to train and update the global models \cite{10.1145/3659097}. As for privacy concerns, federated learning (FL) among distributed edge-cloud computing nodes emerges as an alternative paradigm, enabling collaborative model training without sharing raw data. By keeping user interactions on-device and sharing only aggregated updates, this paradigm offers a principled path to strong privacy guarantees~\cite{GUENDOUZI2023103714}. However, deploying Federated Recommendation Systems (FRS) in a practical edge-cloud computing environment entails its own set of challenges due to intrinsic complexities of distributed scheduling \cite{fogbus2book}, which include heterogeneous hardware settings, non-Independent and Identically Distributed (non-IID) data across devices, communication bottlenecks \cite{9591490}, and the difficulty of balancing data skew in global aggregation especially with multimodal data \cite{10.1145/3625558}. In this context, participant selection among heterogeneous clients plays a critical role in improving the efficiency and robustness of federated learning.     

In FL systems, participant selection strategies are broadly categorized into two approaches: 1) data utility maximization and 2) system utility optimization \cite{LI2024110663}. Data utility maximization strategies employ importance-based ranking mechanisms that leverage various metrics, including gradient norm, local loss \cite{LI2024110663}, Shapley value \cite{10.1145/3580305.3599500}, and Client Performance Reputation (CPR) \cite{10361025} to identify the most valuable participants. These approaches are complemented by clustering-driven methodologies such as hierarchical grouping and K-means clustering to organize clients based on data characteristics. System utility optimization strategies utilize optimization techniques to achieve optimal trade-offs between global model objectives and practical constraints such as bandwidth limitations and computational resources \cite{10197174}. Additionally, Reinforcement Learning (RL) frameworks enable dynamic exploration-exploitation trade-offs during client sampling \cite{WANG2025104250}, while advanced hybrid frameworks integrate multiple selection criteria through a two-stage utility optimization combined with deep reinforcement learning (DRL) methodologies \cite{LI2024110663}. 

Despite theoretical advances in participant selection for FRS, practical deployment remains constrained by the intricate coupling of device heterogeneity and diverse data characteristics. Existing literature can be broadly categorized into three streams: 1) security-centric approaches utilizing differential privacy and clustering to mitigate malicious behavior \cite{10562343}; 2) fairness-oriented studies ensuring equitable participation \cite{WANG2024110234}; and 3) heterogeneity-aware solutions addressing cross-silo integration and cold-start issues \cite{10349125}. However, these approaches typically address constraints in isolation or assume static environments. They struggle to simultaneously model the dynamic trade-off between multimodal data utility and fluctuating system costs (e.g., latency, availability), leading to sub-optimal convergence and resource inefficiency in dynamic, real-world scenarios.

To tackle these interconnected challenges, we formulate the FRS participant selection problem as a normalized utility cost addressing the model quality and system efficiency. Specifically, we propose a dynamic Multi-Armed Bandit (MAB)-based framework to solve this problem in realistic, time-varying edge–cloud environments. Unlike static heuristics, our approach dynamically addresses the aforementioned limitations by treating client selection as an online decision-making process under uncertainty. By modeling clients as arms, the MAB strategy orchestrates a principled balance between \textit{exploration}—discovering under-sampled clients to mitigate data skew and cold-start problems—and \textit{exploitation}—selecting clients with proven high utility to ensure system efficiency. This adaptive mechanism allows the system to robustly handle volatile hardware capabilities and communication bottlenecks while maximizing the global model performance.

The main contributions of this paper are summarized below:
\begin{enumerate}
  \item We propose a unified utility optimization formulation that integrates historical Client Performance Reputation (CPR), system efficiency, and data quality into the RL reward, enabling dynamic trade‐offs between accuracy and efficiency.
  \item We propose a MAB-based selection algorithm that learns exploration–exploitation policies in dynamic federated environments, improving training efficiency and model performance under client heterogeneity.
  \item We implement and evaluate our method in practical edge--cloud FL settings against the baselines, demonstrating up to 50\% acceleration in time-to-target convergence across eight data distributions. Crucially, this efficiency is achieved while maintaining competitive or superior performance in Area Under the Receiver Operating Characteristic Curve (AUC), Normalized Discounted Cumulative Gain at 50 (NDCG@50), and Recall at 50 (Recall@50) compared to standard baselines of FedAvg \cite{mcmahan2017communication} and RPFL~\cite{10562343}.
\end{enumerate}

The remainder of this paper is organized as follows. Section~II reviews related work on multimodal RS, FL, and selection strategies. Section~III presents the system model and formalizes the selection problem. Section~IV details the RL formulation and the proposed MAB algorithm. Section~V describes the evaluation setup and reports experimental results on the MovieLens-100K dataset. Finally, Section~VI concludes the paper and outlines future research directions.

\section{Background and Related Work}

This section provides an overview of three topics: Multimodal RS, FL, and Participant Selection. Finally, related works are described in this section.

\subsection{Multimodal Recommendation System}

 Model-based collaborative filtering (CF) methods project users and items into a shared latent space, where interactions are predicted via similarity functions. Early successes include MF techniques, which minimize the reconstruction error of the user–item matrix, and probabilistic extensions such as Probabilistic Matrix Factorization (PMF)  \cite{10925927}. To better capture non-linear interactions, Neural Collaborative Filtering (NCF) replaces the inner product with multi-layer perceptrons \cite{10.1145/3038912.3052569}. However, these methods often underperform in sparse regimes or cold-start scenarios, since they rely exclusively on interaction histories. To alleviate sparsity, auxiliary content such as textual reviews and visual features have been fused with CF. Visual Bayesian Personalized Ranking (VBPR) injects pretrained image embeddings into the ranking objective \cite{10.5555/3015812.3015834}. However, fusing heterogeneous modalities remains challenging due to distributional mismatch and complementary information gaps. Disentangled multimodal representation learning (DMRL) disentangles fine-grained text and image factors to better model item characteristics \cite{10.1109/TMM.2022.3217449}.

\begin{figure}[htbp]
  \centering
  \includegraphics[width=0.5\textwidth]{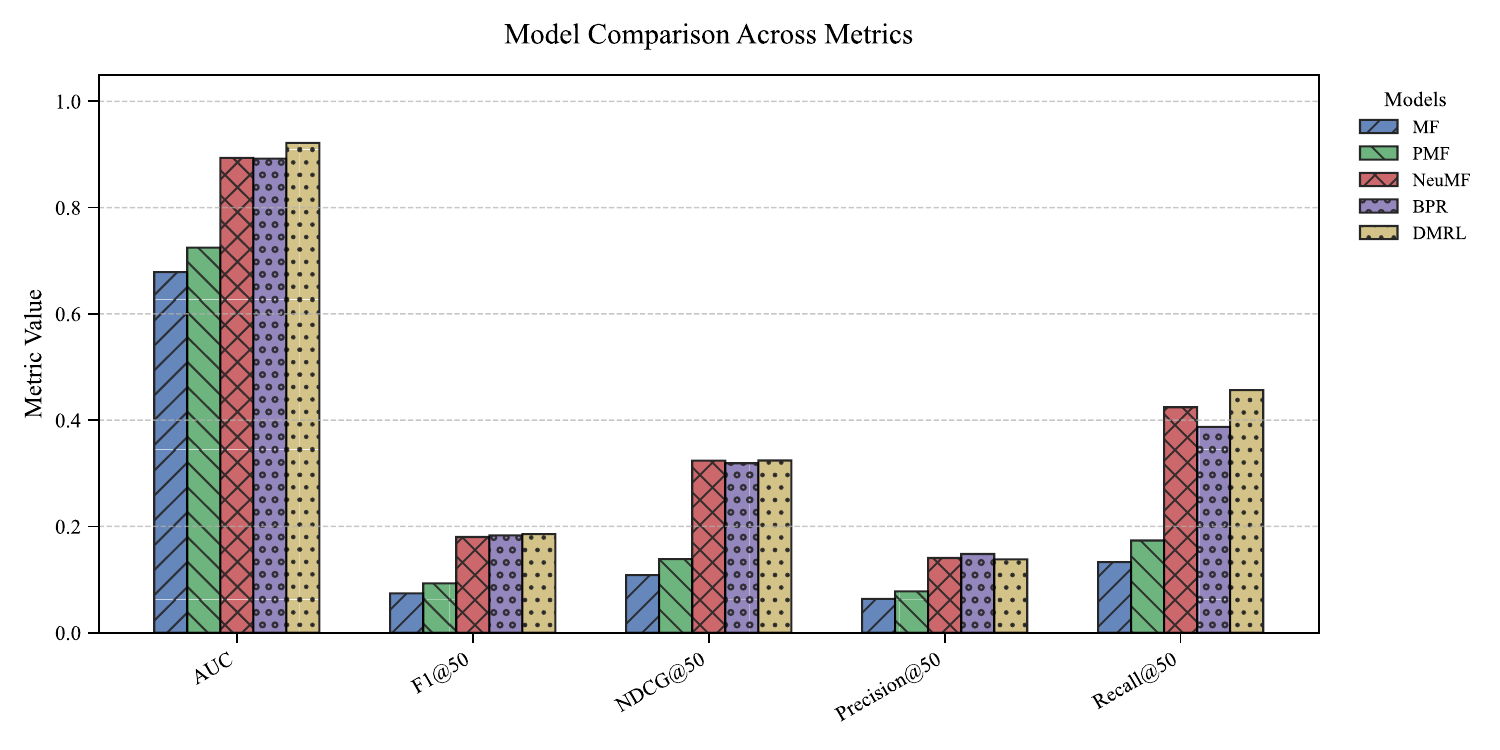}
  \caption{Model Evaluation over Recommendation Metrics}
  \label{fig:model-timing}
\end{figure}

As shown in Figure~\ref{fig:model-timing}, we evaluate the performance of centralized RS models on the widely used MovieLens-100K benchmark \cite{10.1145/2827872} to justify the selection of a multimodal RS as the base model in our FL framework. Model effectiveness is assessed using Precision, Recall, NDCG, F1-score, and AUC. As illustrated in the figure, the multimodal RS model consistently outperforms other models across all evaluation metrics. In particular, the DMRL-based model achieves an AUC of 0.9219, an F1@50 of 0.1856, and an NDCG@50 of 0.3241. This superior performance can be attributed to the ability of multimodal RS models to learn richer and more complementary feature representations by jointly exploiting heterogeneous data sources, including collaborative signals, textual metadata, and visual item features.

\subsection{Federated Learning}

FL is a decentralized training paradigm in which a central server orchestrates model updates across a large number of clients that each retain their data locally. In every communication round, the server sends the current global model parameters to a selected group of clients \cite{10.1145/3625558}. These clients then perform several epochs of local optimization on their own datasets and return only their model updates or gradients. When the server aggregates these updates, usually by weighting them according to the size of each client’s dataset, it produces an improved global model without ever accessing raw data. This iterative process continues until the model converges, enabling collaborative learning while preserving user privacy \cite{GUENDOUZI2023103714}. Deploying FL at scale introduces several practical challenges. First, non-IID distributed data across clients can bias local updates and slow convergence. Second, constrained bandwidth and unreliable network connections can delay or cause training dropouts. Third, the heterogeneous distribution of computational resources gives rise to straggler problems and induces training delays \cite{10.1145/3625558}. To mitigate these issues, adaptive aggregation methods can be used to adjust each client’s contribution based on criteria such as data volume and the recency of updates \cite{Wen2023}. Moreover, intelligent participant selection is essential: by choosing clients with diverse data profiles or sufficient computational resources, the system can improve training efficiency, maintain representation fairness, and ultimately enhance global model performance \cite{LI2024110663}.

\subsection{Related Work}

Within the specific context of FRS, participant selection refers to the process by which the server chooses a subset of clients to contribute updates at each communication round. Formally, let $\mathcal{C}$ be the full set of $K$ clients and $\mathcal{S}_t \subseteq \mathcal{C}$ the clients selected at round $t$. The choice of $\mathcal{S}_t$ directly impacts convergence speed, model quality, and fairness across heterogeneous clients \cite{10542323}. 

\begin{table}[!t]
  \centering
  \caption{A Comparative Analysis of Participant Selection Approaches in FRS}
  \label{tab:table_1}
  \footnotesize
  \setlength{\tabcolsep}{2pt} 
  \renewcommand{\arraystretch}{0.9}
  
  
  \begin{tabularx}{\columnwidth}{@{}l*{5}{Y}@{}}
    \toprule
    Method
      & \multicolumn{5}{c}{Data/System Heterogeneity} \\
    \cmidrule(l){2-6}
      & \makecell{Mode}
      & \makecell{\scriptsize Data\\\scriptsize Distribution} 
      & \makecell{\scriptsize Comm.\\\scriptsize Delay} 
      & \makecell{\scriptsize Compu.\\\scriptsize Capacity} 
      & \makecell{CPR*} \\ 
    \midrule
    RPFL \cite{10562343}              & Simulated & \cmark & \xmark & \xmark & \xmark \\
    HeteFedRec \cite{10598074}        & Simulated & \cmark & \xmark & \cmark & \xmark \\
    AeroRec \cite{10621240}           & Simulated & \cmark & \cmark & \xmark & \xmark \\
    FedRAP \cite{li2024federated}     & Simulated & \cmark & \cmark & \xmark & \xmark \\
    FedConPE \cite{10.24963/ijcai.2024/501} & Simulated & \cmark & \cmark & \cmark & \xmark \\
    CoFedRec \cite{10.1145/3589334.3645626} & Simulated & \cmark & \xmark & \xmark & \xmark \\
    PriFedRec \cite{10476059}         & Simulated & \cmark & \xmark & \xmark & \xmark \\
    FedAvg \cite{mcmahan2017communication} & Simulated & \xmark & \xmark & \xmark & \xmark \\
    \midrule
    Our Method                        & Practical & \cmark & \cmark & \cmark & \cmark \\
    \bottomrule
  \end{tabularx}
  
  \vspace{1ex} 
  \raggedright \scriptsize * CPR: Client Performance Reputation. 

  \vspace{-5ex} 
\end{table}

In recent studies, Yuan \textit{et al.}\  \cite{10598074} proposed HeteFedRec, a federated framework supporting heterogeneous model selection and aggregation mechanisms. Xia \textit{et al.}\ \cite{10621240} introduced AeroRec, an on-device FRS model employing federated self-supervised distillation and selection to boost lightweight-model performance. Feng \textit{et al.}\ \cite{10562343} designed the privacy-preserving multimodal RS with clustering-based participant selection to optimize statistical resilience. Li  \textit{et al.}  \cite{li2024federated} presented FedRAP, which learns and selects additive global and personalized item embeddings with sparse regularization for communication efficiency. Li \textit{et al.}  \cite{10.24963/ijcai.2024/501} proposed FedConPE, a federated conversational bandit algorithm optimizing communication and regret. He \textit{et al.}\ \cite{10.1145/3589334.3645626} proposed CoFedRec, a co-clustering mechanism grouping clients and items based on network structure and embedding similarity. Ting  \textit{et al.} \cite{10476059} introduced PriFedRec, which integrates homomorphic encryption for privacy and evaluates client availability and communication delay when selecting participants, demonstrating reduced convergence time.

In summary, common participant selection strategies can be classified into the following categories:
\begin{enumerate}
  \item \textbf{Random Selection:}  
    Each round, a fixed number $S$ of clients is chosen uniformly at random. This simple approach guarantees unbiased participation in expectation but can repeatedly choose clients with poor connectivity or small data, delaying the convergence in the global model \cite{mcmahan2017communication}.
  \item \textbf{Data-Aware Selection:}  
    Clients whose local gradients have higher variance or whose data distributions differ more from the global distribution may be sampled more frequently to accelerate convergence \cite{10562343,10598074,10621240,li2024federated,10.24963/ijcai.2024/501,10.1145/3589334.3645626,10476059, 273723}. However, these methods typically require additional computation and communication overhead to estimate data heterogeneity or gradient statistics, which may not be readily available or accurately measurable in practice. Moreover, overemphasizing highly heterogeneous clients can lead to biased updates, increased instability during training, and potential fairness concerns, especially when certain clients are persistently oversampled.
\item \textbf{Resource-Aware Selection:}  
    Client sampling is guided by availability, communication bandwidth, and computational capacity. Classic general-purpose client selection frameworks, such as Oort \cite{273723}, have pioneered the optimization of time-to-accuracy by profiling client data and system utility. By prioritizing clients, this approach mitigates straggler effects and reduces overall training latency \cite{10598074,10621240,li2024federated,10.24963/ijcai.2024/501}. Despite their efficiency advantages, these strategies may systematically exclude resource-constrained clients, resulting in under-utilization of valuable data and reduced model generalization. In addition, prioritizing well-resourced clients can exacerbate data imbalance and introduce selection bias, ultimately compromising model fairness and robustness.

\end{enumerate}

\section{System Model and Problem Formulation}

In this section, we first define a multimodal RS model with user ID, text, and visual embeddings, factor‐wise attention, and a unified scoring function. We then introduce a client utility optimization formulation with CPR, system quality, and data quality, and incorporate per‐round latency into a normalized time model. Finally, we formalize participant selection as an optimization that maximizes utility under delay constraints.

\subsection{Multimodal Recommendation Backbone}

We adopt a multimodal recommendation model as the backbone architecture, which is later integrated into the FL framework. The design follows the paradigm of DMRL~\cite{10.1109/TMM.2022.3217449} and models user--item interactions from multiple modalities.

Let $\mathcal{U}$ denote the set of users, indexed by $u \in \mathcal{U}$, and $\mathcal{I}$ denote the set of items, indexed by $i \in \mathcal{I}$.  
For each user $u$, a learnable user ID embedding $\mathbf{p}_u \in \mathbb{R}^d$ is maintained.  
For each item $i$, we consider three modalities:
\begin{itemize}
    \item ID modality: item ID embedding $\mathbf{v}_i \in \mathbb{R}^d$;
    \item Text modality: raw textual feature vector $\mathbf{t}_i \in \mathbb{R}^{d_T}$;
    \item Visual modality: raw image feature vector $\mathbf{x}_i \in \mathbb{R}^{d_V}$.
\end{itemize}

\subsubsection{Modality Projection and Factor-Wise Decomposition}

To project heterogeneous modality features into a shared latent space, we employ two independent two-layer multilayer perceptrons (MLP) for text and image modalities:
\begin{align}
\mathbf{t}_i' &= \mathrm{MLP}_T(\mathbf{t}_i) \in \mathbb{R}^d, \notag \\
\mathbf{x}_i' &= \mathrm{MLP}_V(\mathbf{x}_i) \in \mathbb{R}^d.
\end{align}

Let $\mathcal{M} = \{\mathrm{ID}, T, V\}$ denote the set of item modalities. The modality-specific item representations are defined as
\begin{equation}
\mathbf{e}_i^{(\mathrm{ID})} = \mathbf{v}_i, \quad
\mathbf{e}_i^{(T)} = \mathbf{t}_i', \quad
\mathbf{e}_i^{(V)} = \mathbf{x}_i'.
\end{equation}

To enable disentangled representation learning, each $d$-dimensional embedding is evenly partitioned into $F$ latent factors, such that $d = F \cdot d_f$. Accordingly, $\mathbf{p}_u^{(f)} \in \mathbb{R}^{d_f}$ and $\mathbf{e}_i^{(m,f)} \in \mathbb{R}^{d_f}$ denote the $f$-th factor block of the user embedding and the item embedding under modality $m \in \mathcal{M}$, respectively.

For a given user--item pair $(u,i)$ and factor $f \in \{1,\dots,F\}$, the factor-wise multimodal interaction vector is constructed as
\begin{equation}
\mathbf{h}_{u,i}^{(f)} =
\bigl[
\mathbf{p}_u^{(f)};
\mathbf{v}_i^{(f)};
\mathbf{t}_i^{\prime(f)};
\mathbf{x}_i^{\prime(f)}
\bigr]
\in \mathbb{R}^{4d_f}.
\end{equation}

\subsubsection{Factor-Wise Multimodal Attention and Prediction}

To adaptively fuse different modalities at each latent factor, we employ a factor-wise attention mechanism. Specifically, an attention network $\mathrm{MLP}_{\mathrm{Attn}}$ takes $\mathbf{h}_{u,i}^{(f)}$ as input and outputs normalized modality attention weights:
\begin{equation}
\boldsymbol{\alpha}_{u,i}^{(f)} =
\mathrm{Softmax}
\bigl(
\mathrm{MLP}_{\mathrm{Attn}}(\mathbf{h}_{u,i}^{(f)})
\bigr),
\end{equation}
where $\boldsymbol{\alpha}_{u,i}^{(f)} = \{\alpha_{u,i,m}^{(f)} \mid m \in \mathcal{M}\}$ and $\sum_{m \in \mathcal{M}} \alpha_{u,i,m}^{(f)} = 1$.

The factor-specific interaction score between user $u$ and item $i$ is computed as
\begin{equation}
s_{u,i}^{(f)} =
\sum_{m \in \mathcal{M}}
\alpha_{u,i,m}^{(f)}
\cdot
\mathrm{GELU}
\bigl(
\langle
\mathbf{p}_u^{(f)},
\mathbf{e}_i^{(m,f)}
\rangle
\bigr),
\end{equation}
where $\langle \cdot, \cdot \rangle$ denotes the inner product, and $\mathrm{GELU}(\cdot)$ denotes the Gaussian Error Linear Unit, a smooth non-linear activation function that scales the input by the probability of being activated under a standard Gaussian distribution. 

Finally, the predicted preference score is obtained by aggregating all factor-wise scores:
\begin{equation}
\hat{r}_{u,i} = \sum_{f=1}^{F} s_{u,i}^{(f)}.
\end{equation}

\subsection{Problem Formulation}

We consider an FL system with a set of client nodes $\mathcal{K}$. In each communication round $t$, the server selects a subset $\mathcal{S}_t \subseteq \mathcal{K}$ to maximize the collective utility under a round-latency constraint. To quantify each client’s utility, we adopt a modular scoring scheme that evaluates client $k$ from three perspectives: (i) a customized CPR, (ii) an updated relevance score, and (iii) a data quality metric. These components are aggregated into an overall contribution score, which is then balanced against the expected round time to determine the selection $\mathcal{S}_t$.

\subsubsection{Customized Client Performance Reputation}

We first quantify each client’s historical contribution to the global model performance via a customized performance reputation. Let $Q_{t}$ denote the validation accuracy (or equivalently, the validation performance metric) of the global model after aggregating the update from client $k$ in round t. Let $Q_{t-1}$ be the validation accuracy of the global model at the end of round $t\!-\!1$. We define the marginal accuracy gain attributed to client $k$ in round $t$ as
\begin{equation}
\Delta_{t,k} = Q_{t} - Q_{t-1}.
\end{equation}
Then, the reputation score of client $k$ is updated using exponential smoothing:
\begin{equation}
R_{t,k} = \gamma \Delta_{t,k} + (1-\gamma)R_{t-1,k},
\label{eq:custom_rep}
\end{equation}
where $\gamma \in [0,1]$ controls the trade-off between the most recent gain and historical reputation. We initialize $R_{0,k}=0$ to ensure a neutral starting point. This design rewards clients that consistently improve the global model while attenuating the impact of occasional noisy updates, resulting in a stable yet responsive performance indicator.

\subsubsection{Update Relevance}
To complement the customized client performance reputation, we further evaluate the round-wise relevance of each client update with respect to the current global model.
Let $t$ denote the communication round index and let $\mathbf{w}_t \in \mathbb{R}^{J}$ be the vectorized global model parameters broadcast by the server at the beginning of round $t$, where $J$ is the total number of trainable parameters.
For each client $k \in \mathcal{K}$, let $\mathbf{w}_{t,k} \in \mathbb{R}^{J}$ be the locally trained model parameters returned by client $k$ after performing local optimization initialized from $\mathbf{w}_t$.

We quantify the normalized parameter divergence between the local and global models using the $\ell_1$ norm:
\begin{equation}
  \delta_{t,k}
  = \frac{1}{J}\left\|\mathbf{w}_{t,k}-\mathbf{w}_t\right\|_1
  = \frac{1}{J}\sum_{j=1}^{J}\left|w_{t,k,j}-w_{t,j}\right|.
  \label{eq:dev}
\end{equation}

Let $Q_t$ denote the validation performance of the global model after completing round $t$, and let $Q_{t-1}$ be the validation performance at the end of round $t\!-\!1$.
Based on whether the global model improves in round $t$, we define the update relevance score of client $k$ as
\begin{equation}
  U_{t,k} =
  \begin{cases}
    \exp\!\left(-\delta_{t,k}\right), & \text{if } Q_t > Q_{t-1},\\[4pt]
    1 - \exp\!\left(-\delta_{t,k}\right), & \text{otherwise,}
  \end{cases}
  \label{eq:update_rel}
\end{equation}
where $U_{t,k}\in(0,1)$.
When the global model improves, a smaller divergence $\delta_{t,k}$ yields a larger $U_{t,k}$, encouraging updates that are directionally consistent with the current global trajectory.
When the global performance stagnates or degrades, the mapping is reversed to down-weight updates that remain overly close to $\mathbf{w}_t$, thereby promoting exploratory updates that deviate more substantially.

\subsubsection{Data Quality Measure}
We next characterize the utility of each client’s local data.
Let $\mathcal{B}_k$ denote the local sample set at client $k$ and $|\mathcal{B}_k|$ its cardinality.
Let $\ell_{t,k,n}$ be the training loss of sample $n\in\mathcal{B}_k$ evaluated under the local model in round $t$.
We define a data quality score as
\begin{equation}
  D_{t,k}
  = |\mathcal{B}_k|\,
    \sqrt{\frac{1}{|\mathcal{B}_k|}\sum_{n\in\mathcal{B}_k}\ell_{t,k,n}^2}\,,
  \label{eq:data_quality}
\end{equation}
which increases with both the data volume and the aggregate loss magnitude, reflecting the potential informativeness of underfit, hard, or diverse samples.

To make $D_{t,k}$ comparable across heterogeneous clients, we apply Min--Max normalization over clients:
\begin{equation}
  \widetilde{D}_{t,k}
  = \frac{D_{t,k}-D_t^{\min}}{D_t^{\max}-D_t^{\min}+\varepsilon}
\end{equation}
\begin{equation}
  D_t^{\min}=\min_{j\in\mathcal{K}} D_{t,j},\;
  D_t^{\max}=\max_{j\in\mathcal{K}} D_{t,j},
  \label{eq:data_quality_norm}
\end{equation}
where $\varepsilon>0$ is a small constant to avoid division by zero, and thus $\widetilde{D}_{t,k}\in[0,1]$.

\subsubsection{Aggregate Contribution Score}
We aggregate the three signals (update relevance, CPR, and data quality) into an overall contribution score.
Recall that $R_{t,k}$ denotes the customized client performance reputation defined in~\eqref{eq:custom_rep}.
We define client $k$'s contribution score in round $t$ as
\begin{equation}
  ACS_{t,k}
  = \lambda\,\bigl(U_{t,k}\,R_{t,k}\bigr) + \mu\,\widetilde{D}_{t,k},
  \label{eq:aggregate_score}
\end{equation}
where $\lambda,\mu>0$ control the trade-off between the relevance--reputation term and the normalized data quality.

\subsubsection{Per-Client Latency and Normalized Time Cost}
For client $k$, let $N_k = |\mathcal{B}_k|$ denote its local workload (number of samples),
$v_k$ its computation throughput (samples/s), $Z$ the communicated model size in bits,
and $b_k$ the available bandwidth (bits/s). We estimate the per-client latency as
\begin{equation}
  T_{t,k} = \frac{N_k}{v_k} + \frac{Z}{b_k}.
  \label{eq:client_time}
\end{equation}
To align time cost with utility scores, we normalize the latency by a system-level time budget $T_{\mathrm{semi}}>0$:
\begin{equation}
  \widetilde{T}_{t,k} = \frac{T_{t,k}}{T_{\mathrm{semi}}}.
  \label{eq:time_norm_client}
\end{equation}

\subsubsection{Optimal Selection via Additive Utility--Cost Trade-off}
At each round $t$, the server selects a subset $\mathcal{S}_t \subseteq \mathcal{K}$ with a fixed budget $|\mathcal{S}_t|=M$ by maximizing an additive utility--cost objective:
\begin{equation}
  \max_{\mathcal{S}_t\subseteq\mathcal{K},\,|\mathcal{S}_t|=M}
  \sum_{k\in\mathcal{S}_t}\left( ACS_{t,k} - \kappa\,\widetilde{T}_{t,k} \right),
  \label{eq:selection_opt_additive}
\end{equation}
where $\kappa>0$ controls the trade-off between contribution and latency.
This additive formulation is consistent with top-$M$ ranking and enables efficient online learning with feedback.

\begin{figure}
    \centering
    \includegraphics[width=1.0\columnwidth]{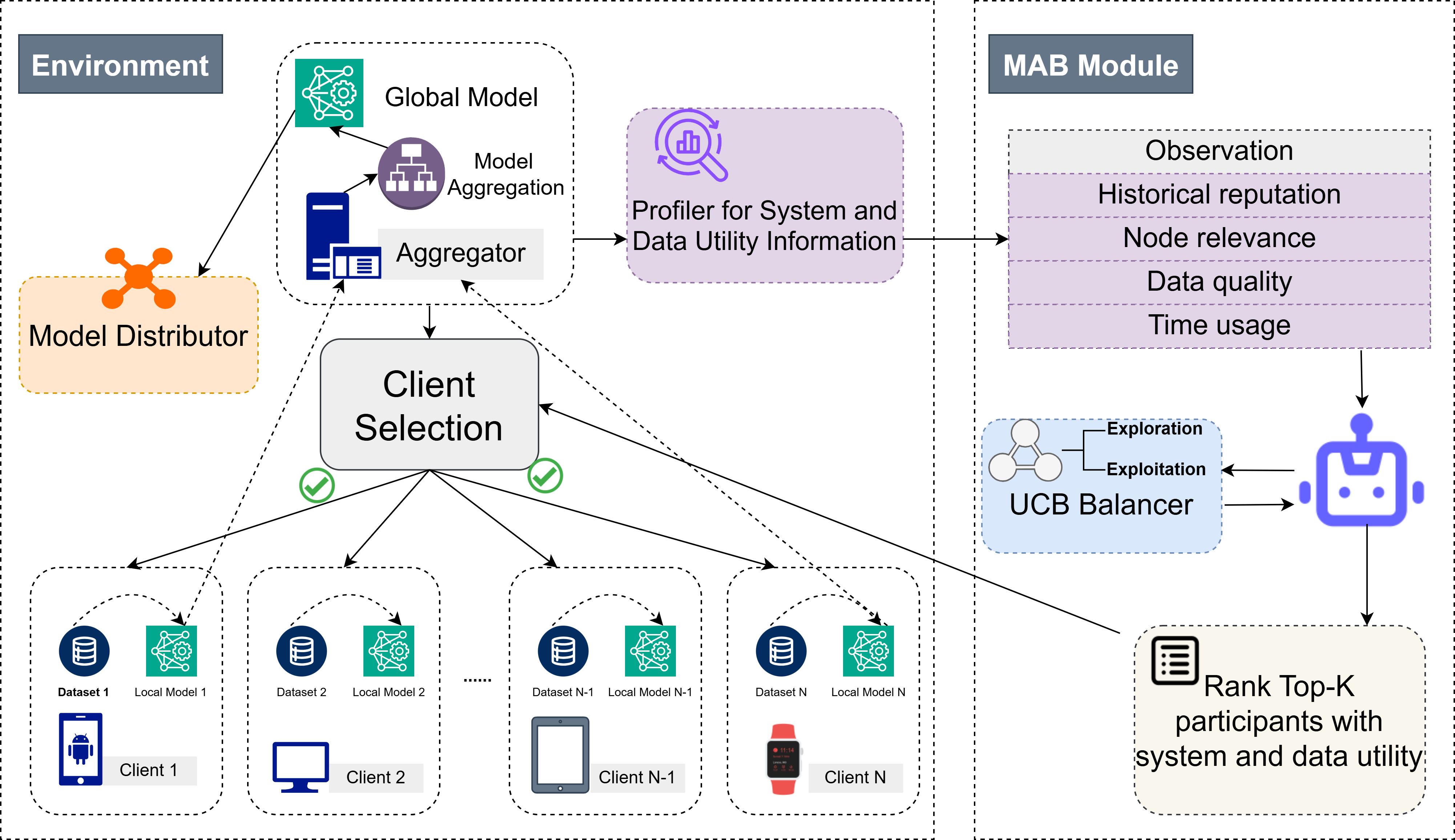}
    \caption{Framework of the Upper Confidence Bound-based Multi-Armed Bandit for Federated Recommendation System participant selection.}
    \label{fig:MAB-Framework}
\end{figure}

\section{Proposed MAB-Driven Participant Selection}

In this section, we formulate the participant selection problem in FL as a sequential decision-making task using the MAB framework. We present the detailed execution flow in Algorithm~\ref{alg:mab-selection}, where the agent iteratively selects clients to maximize global model performance while adhering to resource constraints. The proposed method balances exploration and exploitation through the Upper Confidence Bound (UCB) mechanism.

In our setting, the agent is the coordinator who repeatedly decides which clients to schedule in each communication round.
The arms correspond to candidate clients, where each client $k \in \mathcal{K}$ is treated as one arm that can be pulled by selecting it to participate in a round. At round $t$, the agent selects a subset of clients $\mathcal{S}_t \subseteq \mathcal{K}$ with a fixed budget $|\mathcal{S}_t|=M$. Upon pulling arm $k$ or selecting client $k$, the server observes a stochastic feedback signal reflecting both the client’s contribution and its system cost.

The execution of our proposed framework proceeds in three key phases: Initialization, UCB-based Selection, and Reward Update. We describe each phase below, explicitly referencing the corresponding steps in Algorithm~\ref{alg:mab-selection}.

\subsubsection{Initialization (Line~\ref{line:init})}
At the beginning of the FL process, the agent initializes the empirical reward estimate $\hat{\mu}_k$ and the selection count $n_k$ for every client $k \in \mathcal{K}$ to zero. This cold-start state implies that initially, all clients have an equal probability of being explored based on the exploration bonus defined subsequently.

\subsubsection{UCB Index Construction (Lines~\ref{line:loop_start}--\ref{line:ucb_calc})}
For each communication round $t$, the agent must evaluate the potential value of each client. We calculate the UCB index $I_k(t)$ for every client $k$. This index serves as a surrogate for the client's expected future contribution and is defined as:
\begin{equation}
    I_k(t) = \hat{\mu}_k + \rho \sqrt{\frac{\ln t}{n_k + 1}},
    \label{eq:ucb_index}
\end{equation}
where $\hat{\mu}_k$ represents the exploitation term (historical performance), and the second term represents exploration. The parameter $\rho$ controls the degree of exploration, and the term $\sqrt{\frac{\ln t}{n_k + 1}}$ naturally decays as a client is selected more frequently, encouraging the system to explore under-sampled clients.

\subsubsection{Candidate Selection (Line~\ref{line:selection})}
Once the indices are computed, the participant selection transforms from a combinatorial optimization problem into a ranking operation. The agent selects the subset of clients $\mathcal{S}_t$ by picking the top-$M$ candidates with the highest $I_k(t)$ values:
\begin{equation}
  \mathcal{S}_t
  = \arg\max_{\mathcal{C}\subseteq\mathcal{K},\,|\mathcal{C}|=M}
  \sum_{k\in\mathcal{C}} I_k(t).
  \label{eq:candidate_selection}
\end{equation}

\subsubsection{Execution and Observation (Line~\ref{line:execute})}
The selected clients $\mathcal{S}_t$ then perform local training and transmit their updates. During this phase, the server observes the system feedback, specifically the contribution score and the resource consumption metrics.

\subsubsection{Reward Calculation and Policy Update (Lines~\ref{line:update_start}--\ref{line:update_end})}
Finally, the algorithm updates its reward estimates based on the observed feedback. We compute a composite reward $r_{t,k}$ for each participating client that penalizes high latency, and is derived from Eq.~\ref{eq:selection_opt_additive}: 
\begin{equation}
    r_{t,k} = ACS_{t,k} - \kappa \cdot \tilde{T}_{t,k},
    \label{eq:reward_calc}
\end{equation}
where $\kappa$ is a penalty coefficient balancing utility and cost. The agent then updates the empirical mean $\hat{\mu}_k$ and increments the selection count $n_k$ using an incremental average update rule, where $n_k$ denotes the updated number of selections of client $k$ up to the current round.

\begin{equation}
    \hat{\mu}_k \leftarrow \hat{\mu}_k + \frac{r_{t,k} - \hat{\mu}_k}{n_k}.
    \label{eq:miuk}
\end{equation}

This update corresponds to maintaining the empirical mean of observed rewards, which serves as a consistent estimate of a client’s average contribution over time, and enables the policy to gradually adapt to evolving client behavior in dynamic federated environments.

\subsubsection{Discussion on Cold-Start}
At the onset of training, the absence of historical data ($n_k = 0$) results in identical UCB indices for all clients, and the algorithm initiates with random sampling. As communication rounds progress and system feedback is continuously gathered, the policy smoothly transitions to utility-driven exploitation, guided by the rapid updates to the empirical reward estimates ($\hat{\mu}_k$).

\begin{algorithm}[t]
\caption{MAB-Driven Participant Selection Strategy}
\label{alg:mab-selection}
\begin{algorithmic}[1]
\REQUIRE Client set $\mathcal{K}$, budget $M$, exploration parameter $\rho$, penalty $\kappa$
\ENSURE Selected subset $\mathcal{S}_t$ for each round

\STATE \textbf{Initialize:} $\hat{\mu}_k \leftarrow 0$, $n_k \leftarrow 0, \forall k \in \mathcal{K}$ \label{line:init}

\FOR{round $t=1,2,\dots,T$} \label{line:loop_start}
    \FOR{each client $k \in \mathcal{K}$} \label{line:ucb_loop_start}
        \STATE $I_k(t) \leftarrow \hat{\mu}_k + \rho\sqrt{\frac{\ln t}{n_k+1}}$ \label{line:ucb_calc} (Eq.~\ref{eq:ucb_index})

    \ENDFOR \label{line:ucb_loop_end}
    \STATE Select $\mathcal{S}_t \leftarrow$ top-$M$ clients with largest $I_k(t)$ \label{line:selection} (Eq.~\ref{eq:candidate_selection})
    \STATE Execute local training for clients in $\mathcal{S}_t$ and aggregate updates \label{line:execute}

    \FOR{each selected client $k \in \mathcal{S}_t$} \label{line:update_start}
        \STATE Observe $ACS_{t,k}$ and normalized latency $\widetilde{T}_{t,k}$ \label{line:observe}
        \STATE $r_{t,k} \leftarrow ACS_{t,k} - \kappa\widetilde{T}_{t,k}$ \label{line:reward}\quad (Eq.~\ref{eq:reward_calc})
        \STATE $n_k \leftarrow n_k + 1$ \label{line:count}
        \STATE $\hat{\mu}_k \leftarrow \hat{\mu}_k + \frac{r_{t,k}-\hat{\mu}_k}{n_k}$ \label{line:update_mean} \quad (Eq.~\ref{eq:miuk})
    \ENDFOR \label{line:update_end}
\ENDFOR \label{line:loop_end}

\end{algorithmic}
\end{algorithm}

\section{Performance Evaluation}

This section presents our system setup and evaluation of our method in terms of training efficiency and recommendation quality, compared against established FL baselines under various system and data heterogeneity settings. 

\subsection{System Configuration}

To emulate a heterogeneous FL environment, we deployed our experiments across eight clients with varying computational and network capabilities, following settings similar to FLASH-RL, KD-AFRL, and ReinFog \cite{10361025,DBLP:journals/corr/abs-2508-21328,WANG2025104250}. As detailed in Table~\ref{tab:client-specs}, the participating clients exhibit a range of heterogeneous configurations to represent a heterogeneous cloud-edge computing environment, with CPU frequencies of approximately 2245 MHz \cite{DBLP:journals/corr/abs-2508-21328}. The cluster includes two tiers of clients: four high-resource nodes, each equipped with 8 CPU cores and 16 GB of RAM, and four low-resource nodes, each with 2 CPU cores and 4 GB of RAM. The network speed of these clients ranges from 2 to 1600 Mbps \cite{10361025, WANG2025104250}. This setup enables the emulation of real-world device heterogeneity typical in FL deployments.

\begin{table}[htbp]
  \centering
  \caption{Client Device Specifications}
  \label{tab:client-specs}
  \begin{tabular}{l c c c c}
    \hline
    \textbf{Name} & \textbf{CPU Frequency} & \textbf{Cores} & \textbf{RAM} & \textbf{Network Speed} \\
    \hline
    Client 1 & 2245.78 MHz & 8  & 16 GB & 1600 Mbps\\
    Client 2 & 2245.78 MHz & 8  & 16 GB & 1600 Mbps\\
    Client 3 & 2245.78 MHz & 8  & 16 GB & 100 Mbps\\
    Client 4 & 2245.78 MHz & 8  & 16 GB & 100 Mbps\\
    Client 5 & 2245.78 MHz & 2  & 4 GB & 6 Mbps\\
    Client 6 & 2245.78 MHz & 2  & 4 GB & 6 Mbps\\
    Client 7 & 2245.78 MHz & 2  & 4 GB & 2 Mbps\\
    Client 8 & 2245.78 MHz & 2  & 4 GB & 2 Mbps\\
    \hline
  \end{tabular}
  \vspace{1mm}
\end{table}

\subsection{Dataset Configuration}

We conduct all experiments on the MovieLens-100K dataset, which contains 100,000 ratings from 943 users on 1,682 items \cite{10.1145/2827872}.  In addition to numerical ratings, we augmented the dataset with different modalities inspired by RPFL~\cite{10562343}, and each item is accompanied by one poster image and one plot summary, enabling multimodal analysis.

\begin{table}[htbp]
  \centering
  \caption{Statistics of the MovieLens-100K Dataset}
  \label{tab:movielens-stats}
  \begin{tabular}{l c}
    \hline
    \textbf{Statistic}      & \textbf{Value} \\
    \hline
    Number of users         & 943            \\
    Number of items         & 1,682          \\
    Number of images      & 1,682          \\
    Number of plot summaries& 1,682          \\
    Number of interactions  & 100,000        \\
    \hline
  \end{tabular}
\end{table}

\begin{figure*}[htbp]
  \centering
  \includegraphics[width=1.0\textwidth, height=5cm]{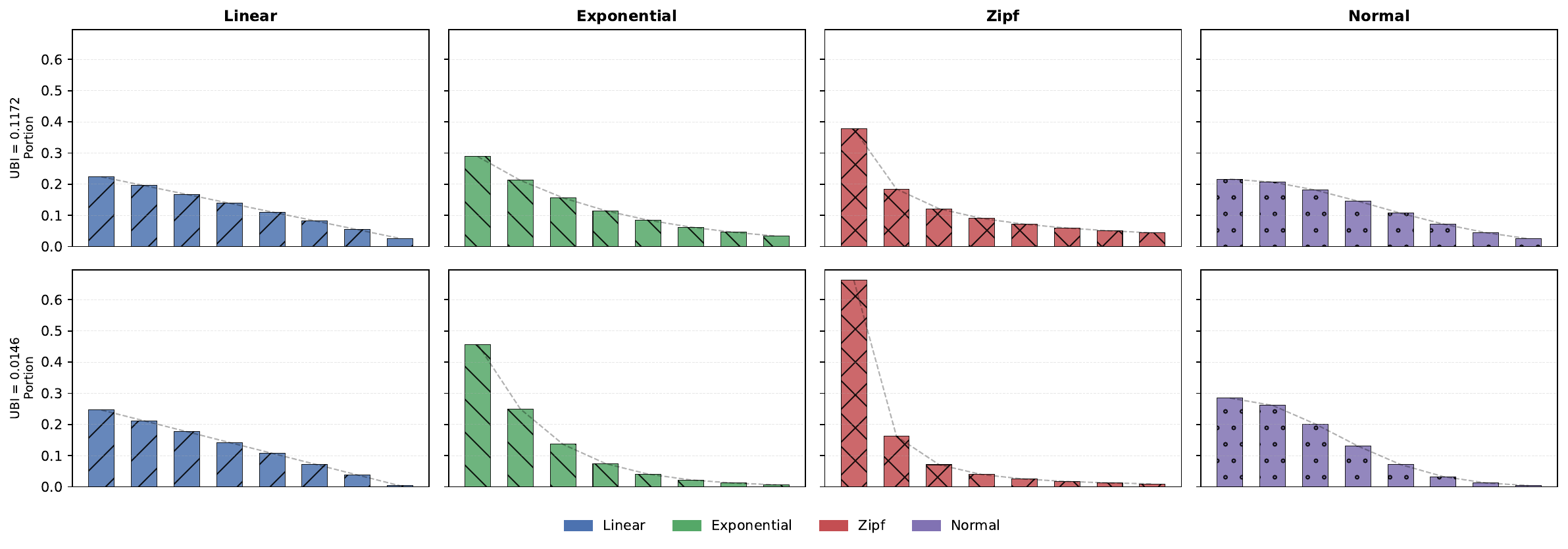}
  \caption{Allocation of data samples to each client under different UBI settings.}
  \label{fig:data-distribution}
\end{figure*}

Inspired by the methodology proposed in RPFL \cite{10562343}, we employed a realistic and heterogeneous data distribution across distributed clients. The MovieLens-100K ratings are partitioned based on a \emph{User Balance Index} (UBI) with linear, exponential, Zipf and normal distributions, which quantifies the degree of sample balance among clients in terms of the number of assigned user interactions.

\begin{equation}
  \mathrm{UBI}
    = \frac{\min\bigl\{\lvert D_{1}\rvert,\ldots,\lvert D_{N}\rvert\bigr\}}
           {\max\bigl\{\lvert D_{1}\rvert,\ldots,\lvert D_{N}\rvert\bigr\}},
  \label{eq:ubi}
\end{equation}
where $\lvert D_i\rvert$ denotes the number of samples assigned to client~$i$.  A UBI close to 1 indicates an almost uniform split, whereas a UBI near 0 reflects a highly skewed allocation. To ensure comparability, the UBI values are set to 0.1172 and 0.0146, consistent with the settings used in RPFL \cite{10562343}. 

We first generate a portion vector $\mathbf{q} = [q_1, \dots, q_N]$ according to one of these strategies—\texttt{exponential}, \texttt{linear}, \texttt{Zipf} and \texttt{normal}—such that
\begin{equation}
  \lvert D_i\rvert \;=\;\bigl\lfloor q_i \cdot \lvert D\rvert \bigr\rfloor,
  \quad i=1,\dots,N,
\end{equation}
where $\lvert D\rvert = 100{,}000$ is the total number of ratings with corresponding modality data.

A visualization of the data distributions under different UBI settings is presented in Figure~\ref{fig:data-distribution}.


\subsection{Federated Learning Baselines}
To evaluate the effectiveness of our proposed method, we compare it against two representative FL baselines:

\textbf{FedAvg} \cite{mcmahan2017communication} is the canonical federated averaging algorithm and serves as a foundational benchmark in FL research. In each communication round, it randomly selects a subset of clients to perform local updates, which are then averaged by the server to update the global model. This approach assumes a homogeneous participation pattern and does not explicitly address statistical heterogeneity or participation constraints.

\textbf{RPFL} \cite{10562343} is a recent method that explicitly accounts for statistical heterogeneity among clients. It employs principal component analysis (PCA) on uploaded gradients to identify clusters of clients with similar data distributions. Based on these clusters, RPFL seeks to maximize the number of participants per round while preserving representativeness in the federated updates, thereby improving both convergence stability and final model performance under non-IID settings.

\subsection{Results and Discussion}

\begin{table*}[htbp]
  \centering

\footnotesize
  \caption{Performance Comparison under Different Variants and UBI Settings}
  \label{tab:comparison}
  \sisetup{
    round-mode=places,
    detect-weight=true, 
    detect-family=true
  }
  \begin{tabular}{%
      l   
      l   
      l   
      S[round-precision=3]   
      S[round-precision=3]   
      S[round-precision=4]   
      S[round-precision=4]   
      S[round-precision=4]   
    }
    \toprule
    Distribution & UBI & Method
      & {Total Training Time (s)}
      & {Time to Target (s)}
      & {AUC}
      & {NDCG@50}
      & {Recall@50} \\
    \midrule
    \multirow{6}{*}{Exponential}
      & \multirow{3}{*}{0.0146}
        & RPFL        & 1000.568 & 227.742 & 0.8387 & 0.2698 & 0.2700 \\
      &                   & FedAvg      &  822.532 & 190.634 & 0.8395 & 0.2699 & 0.2721 \\
      &                   & Our Method  & \bfseries 570.299 & \bfseries 130.408 & \bfseries 0.8398 & \bfseries 0.2713 & \bfseries 0.2796 \\
      \cmidrule{2-8}
      & \multirow{3}{*}{0.1172}
        & RPFL        & 1008.267 & 359.545 & \bfseries 0.8339 & \bfseries 0.2711 & \bfseries 0.2722 \\
      &                   & FedAvg      &  824.615 & 303.324 & 0.8325 & 0.2615 & 0.2626 \\
      &                   & Our Method  & \bfseries 534.102 & \bfseries 181.332 & 0.8326 & 0.2681 & 0.2664 \\
    \midrule
    \multirow{6}{*}{Linear}
      & \multirow{3}{*}{0.0146}
        & RPFL        &  998.350 & 373.864 & \bfseries 0.8354 & \bfseries 0.2674 & \bfseries 0.2747 \\
      &                   & FedAvg      &  815.642 & 323.692 & 0.8331 & 0.2628 & 0.2644 \\
      &                   & Our Method  & \bfseries 540.834 & \bfseries 190.698 & 0.8340 & 0.2638 & 0.2673 \\
      \cmidrule{2-8}
      & \multirow{3}{*}{0.1172}
        & RPFL        & 1003.766 & 398.802 & \bfseries 0.8331 & \bfseries 0.2678 & \bfseries 0.2806 \\
      &                   & FedAvg      &  828.005 & 348.800 & 0.8303 & 0.2551 & 0.2702 \\
      &                   & Our Method  & \bfseries 515.406 & \bfseries 206.960 & 0.8307 & 0.2585 & 0.2715 \\
    \midrule
    \multirow{6}{*}{Zipf}
      & \multirow{3}{*}{0.0146}
        & RPFL        &  965.666 & 161.514 & 0.8354 & 0.2540 & 0.2552 \\
      &                   & FedAvg      &  785.782 & 162.744 & 0.8363 & 0.2469 & \bfseries 0.2569 \\
      &                   & Our Method  & \bfseries 512.372 & \bfseries 107.026 & \bfseries 0.8379 & \bfseries 0.2582 & 0.2521 \\
      \cmidrule{2-8}
      & \multirow{3}{*}{0.1172}
        & RPFL        &  956.053 & 298.017 & 0.8343 & \bfseries 0.2642 & \bfseries 0.2629 \\
      &                   & FedAvg      &  777.520 & 258.180 & \bfseries 0.8346 & 0.2616 & 0.2623 \\
      &                   & Our Method  & \bfseries 488.159 & \bfseries 152.277 & 0.8326 & 0.2574 & 0.2562 \\
    \midrule
    \multirow{6}{*}{Normal}
      & \multirow{3}{*}{0.0146}
        & RPFL        &  935.706 & 272.918 & \bfseries 0.8351 & \bfseries 0.2702 & \bfseries 0.2735 \\
      &                   & FedAvg      &  810.530 & 247.518 & 0.8342 & 0.2702 & 0.2706 \\
      &                   & Our Method  & \bfseries 475.869 & \bfseries 140.096 & 0.8329 & 0.2628 & 0.2653 \\
      \cmidrule{2-8}
      & \multirow{3}{*}{0.1172}
        & RPFL        &  958.779 & 410.818 & 0.8332 & 0.2662 & \bfseries 0.2746 \\
      &                   & FedAvg      &  769.687 & 321.566 & \bfseries 0.8337 & \bfseries 0.2701 & 0.2737 \\
      &                   & Our Method  & \bfseries 435.930 & \bfseries 170.746 & 0.8316 & 0.2599 & 0.2672 \\
    \bottomrule
  \end{tabular}
\end{table*}

We assess efficiency via total wall-clock convergence time and the time required to reach a target AUC of 0.82, following the approach of Lai \textit{et al.} \cite{273723} regarding important metrics. All reported results for each practical experiment represent the average of ten independent runs, to minimize randomness and ensure the reliability and consistency of the findings. For recommendation quality, we employ three complementary metrics \cite{10.1109/TMM.2022.3217449}:
\begin{itemize}
    \item \textbf{AUC}: Measures the global ability to discriminate between relevant and non-relevant items.
    \item \textbf{NDCG@50}: Evaluates the ranking quality of the top-50 items, assigning higher weights to relevant items.
    \item \textbf{Recall@50}: Quantifies the proportion of relevant items retrieved within the top-50 recommendations.
\end{itemize}


Table~\ref{tab:comparison} reports total training time, time to target (reach AUC = 0.82), and final recommendation metrics for FedAvg, RPFL, and our MAB‐based method across eight data‐distribution scenarios. Overall, our approach delivers substantial end‐to‐end speedups while matching or slightly improving recommendation quality.

Across all environments, our approach demonstrates a dominant advantage in computational efficiency. Specifically, in the \textit{Exponential} distribution (UBI 0.0146), our method reduces the total training time by {30.67\%} compared to FedAvg and {43.00\%} compared to RPFL. This efficiency gain is even more pronounced in the \textit{Normal} distribution (UBI 0.1172), where our framework reduces the total training time to 435.930s, representing a reduction of {43.36\%} relative to FedAvg and {54.53\%} relative to RPFL.

\begin{figure*}[htbp]
  \centering
  \includegraphics[width=\linewidth, height=8cm]{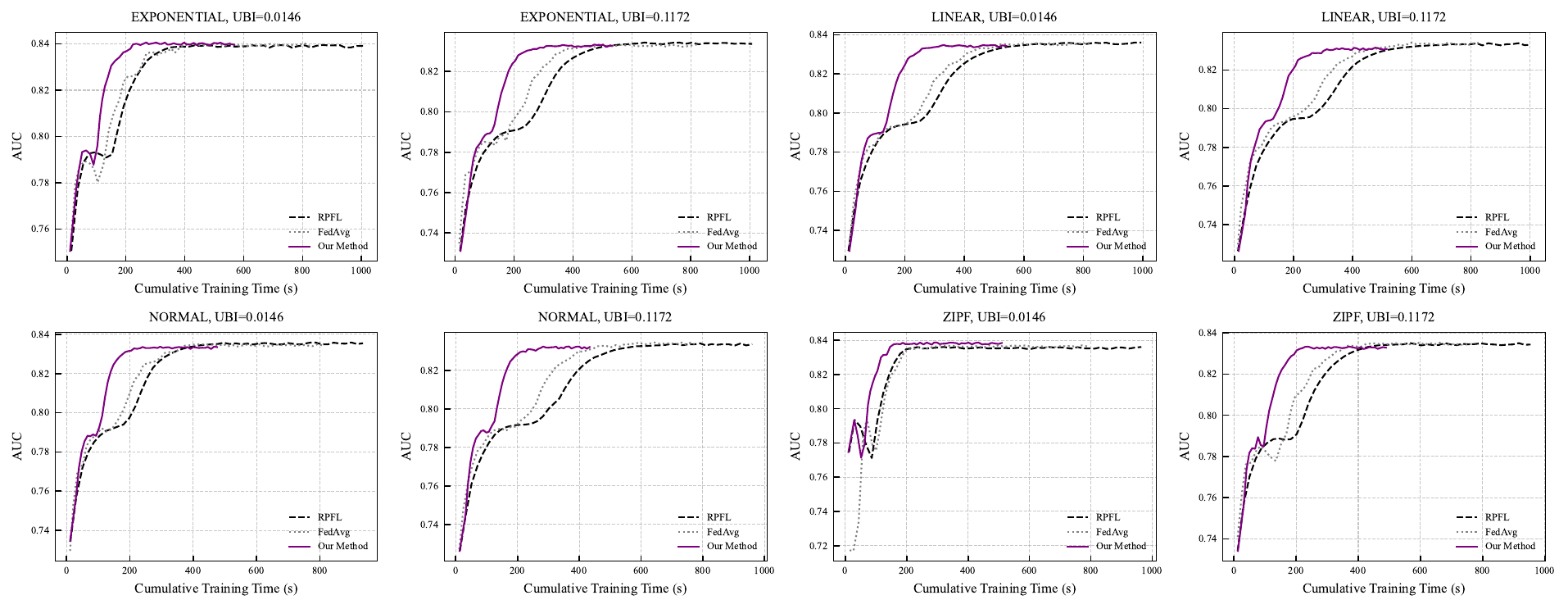}
  \caption{Area Under the Receiver Operating Characteristic Curve (AUC) over Cumulative Training Time for RPFL, FedAvg, and our method under different data partitioning, with different User Balance Index settings.}
  \label{fig:convergence-curves}
\end{figure*}

Furthermore, regarding the convergence speed (Time to Target AUC=0.82), our algorithm consistently identifies the optimal model parameters faster than the baselines. For instance, in the \textit{Zipf} distribution (UBI 0.0146), we achieve the target performance in only 107.026s, which is {34.23\%} faster than FedAvg (162.744s) and {33.73\%} faster than RPFL (161.514s). This validates the effectiveness of our client selection strategy in mitigating stragglers and accelerating global model convergence.

\textbf{Recommendation Quality vs. Latency Trade-off:}
While our primary contribution lies in latency reduction, the results indicate that this speedup does not come at the expense of RS model quality.
\begin{itemize}
    \item \textbf{High-Fidelity Scenarios:} In the \textit{Exponential} distribution (UBI 0.0146), our method not only achieves the fastest convergence but also secures the highest AUC (0.8398), surpassing both RPFL (0.8387) and FedAvg (0.8395). Similarly, in the \textit{Zipf} distribution (UBI 0.0146), we achieve the best performance in AUC ({0.8379}) and NDCG@50 ({0.2582}).
    
    \item \textbf{Robustness under Heterogeneity:} In different scenarios, such as the \textit{Linear} distribution, our method maintains an AUC of 0.8307, which exhibits a negligible degradation of roughly 0.28\% compared to the highest performing baseline (RPFL: 0.8331), yet we deliver a substantial {48.65\%} reduction in total training time.
\end{itemize}

The data highlights a critical characteristic of our MAB-based approach: it strikes a desirable balance between maximizing recommendation accuracy and minimizing system latency. While RPFL and FedAvg occasionally achieve marginally higher Recall or NDCG in specific high-skew settings, the disproportionate cost in training time makes them less viable for real-time applications. Our method, conversely, consistently provides the lowest latency profile with competitive, and often superior, recommendation accuracy. We summarize the main properties of our model as follows.
\par\noindent\textbf{Robustness \& stability under distribution shift.}
Beyond average gains, we observe that our method remains stable across all distribution families and UBI regimes, with consistent time-to-target and closely clustered final AUC/Recall values. This is particularly important when UBI is extremely small (e.g., 0.0146), where FedAvg is more likely to be slowed down by long-tail clients and noisy updates. In contrast, our reward design in Eq.~\eqref{eq:reward_calc} explicitly couples the aggregate contribution $ACS_{t,k}$ with a latency penalty $\kappa\widetilde{T}_{t,k}$, allowing the policy to automatically de-emphasize clients that are either system stragglers or persistent low contributors. As a result, the global training trajectory exhibits less sensitivity to skewed edge populations and maintains reliable convergence behavior even under highly non-uniform data allocations.
\par\noindent\textbf{Why lower UBI yields larger speedups.}
The advantage of intelligent selection becomes more pronounced as the data distribution becomes more imbalanced. When UBI decreases, ``blind aggregation'' tends to admit a larger fraction of low-quality/high-latency updates, making additional participation exhibit diminishing returns. Our UCB ranking in Eq.~\eqref{eq:ucb_index} mitigates this by quickly identifying high-value arms---clients that repeatedly deliver positive utility per unit time---while still preserving controlled exploration via the bonus term $\rho\sqrt{\ln t/(n_k+1)}$. Hence, under extreme skew, selecting \emph{who to train} is more impactful than increasing \emph{how many rounds}, explaining the larger acceleration observed at UBI$=0.0146$.
\par\noindent\textbf{Dual Gains in Accuracy and Multimodal Efficiency.}
Notably, the observed acceleration is not achieved by sacrificing model utility; our approach realizes a simultaneous advancement in both training speed and predictive performance. In several settings (e.g., Exponential with UBI$=0.0146$), our method reduces training time substantially while improving Recall@50 over RPFL. This indicates that $ACS_{t,k}$ (Eq.~\eqref{eq:aggregate_score}) effectively distills generalization-critical information via reputation--relevance and data quality, and the MAB policy functions as an online screening mechanism that filters unstable or low-yield updates. Moreover, because multimodal training amplifies compute and communication bottlenecks, directly incorporating per-client latency (Eq.~\eqref{eq:client_time}) makes the selection resource-intelligent, prioritizing clients that are both fast and informative and thereby improving end-to-end throughput in heterogeneous edge--cloud deployments.

\section{Conclusion and Future Work}

In this paper, we addressed the critical challenges of device heterogeneity and non-IID data in FRS by proposing a novel MAB-driven participant selection framework. Unlike static strategies, our approach dynamically optimizes a composite utility function that integrates historical CPR, data quality, and system efficiency. Empirical results on the multimodal MovieLens-100K dataset demonstrate the effectiveness of our method in realistic edge-cloud environments. Specifically, we achieved a significant acceleration in convergence, reducing time-to-target AUC by 32--50\% and total wall-clock training time by up to 54\% across various skewed data distributions. Crucially, these efficiency gains were realized while maintaining, and in some cases surpassing, the recommendation accuracy of baselines like RPFL and FedAvg.


For future work, we plan to extend this framework in several directions. While our composite utility function provides a foundation, standard MAB algorithms may not fully exploit the intrinsic structural properties of FRS. We will therefore focus on developing customized selection algorithms tailored to recommendation-system dynamics, such as user-item interaction sparsity and collaborative filtering topologies. In addition, benchmarking our framework against general-purpose utility-driven selection schemes, such as Oort, in multimodal settings remains a priority. To further verify generalization, we plan to evaluate the framework on more diverse multimodal datasets and highly skewed synthetic environments. Moreover, beyond bandit-based selection, an important direction for future research is to investigate more expressive DRL and multi-agent reinforcement learning formulations for mitigating server-side bottlenecks in ultra-large-scale deployments. Under such circumstances, distributed DRL and MARL frameworks may provide a more scalable solution for participant selection.

\sloppy
\bibliographystyle{IEEEtran}
\bibliography{refs}
\end{document}